\documentclass[fleqn]{annalen}
\usepackage{graphics}
\usepackage{latexsym}
\usepackage{amssymb}
\begin{document}
%%%%%%%%%%%%%%%%%%%%%%%%%%%%%%%%%%%%%%%%%%%%%%%%%%%%%%%%%%%%%%%%%%%%%%%%%%%%%%
%%%%%%%% the following newcommands will be completed by the publisher %%%%%%%%
%%%%%%%%%%%%%%%%%%%%%%%%%%%%%%%%%%%%%%%%%%%%%%%%%%%%%%%%%%%%%%%%%%%%%%%%%%%%%%
\newcommand{\volume}{8}              %sets current volume,
\newcommand{\xyear}{1999}            %sets year in header
\newcommand{\issue}{5}               %sets current issue,
\newcommand{\recdate}{29 July 1999}  %sets received date,
\newcommand{\revdate}{dd.mm.yyyy}    %sets revised date,     
\newcommand{\revnum}{0}              %number of revisions,
\newcommand{\accdate}{dd.mm.yyyy}    %sets accepted date,
\newcommand{\coeditor}{ue}           %sets (co)editor,
\newcommand{\firstpage}{0}         %first page number,  
\newcommand{\lastpage}{0}          %last page number,
\setcounter{page}{\firstpage}        %sets page counter to first page number 
%%%%%%%%%%%%%%%%%%%%%%%%%%%%%%%%%%%%%%%%%%%%%%%%%%%%%%%%%%%%%%%%%%%%%%%%%%%%%%
%%%%%%%%%%%%%%%%%%%%%%%%%%%%%%%%%%%%%%%%%%%%%%%%%%%%%%%%%%%%%%%%%%%%%%%%%%%%%%
%%%%%%%%%%%%%%%%%% please give up to three keywords here %%%%%%%%%%%%%%%%%%%%%
%%%%%%%%%%%%%%%%%%%%%%%%%%%%%%%%%%%%%%%%%%%%%%%%%%%%%%%%%%%%%%%%%%%%%%%%%%%%%%
\newcommand{\keywords}{frequency dependent conductivity, ac-QHE,
frequency scaling} 
%%%%%%%%%%%%%%%%%%%%%%%%%%%%%%%%%%%%%%%%%%%%%%%%%%%%%%%%%%%%%%%%%%%%%%%%%%%%%%
%%%%%%%%%%%%%%%% please give up to three PACS numbers here %%%%%%%%%%%%%%%%%%%
%%%%%%%%%%%%%%%%%%%%%%%%%%%%%%%%%%%%%%%%%%%%%%%%%%%%%%%%%%%%%%%%%%%%%%%%%%%%%%
\newcommand{\PACS}{71.30.+h, 73.40.Hm, 71.50.+t, 71.55.Jv} 
%%%%%%%%%%%%%%%%%%%%%%%%%%%%%%%%%%%%%%%%%%%%%%%%%%%%%%%%%%%%%%%%%%%%%%%%%%%%%%
%% please enter (First) Author (et al.) and short version of the title here %%
%%%%%%%%%%%% must not exceed 80 characters in length together %%%%%%%%%%%%%%%%
%%%%%%%%%%%%%%%%%%%%%%%%%%%%%%%%%%%%%%%%%%%%%%%%%%%%%%%%%%%%%%%%%%%%%%%%%%%%%%
\newcommand{\shorttitle}{A. B\"aker and L. Schweitzer, Frequency
dependent conductivity in the integer QHE} 
%% sets the header on oddpage
%%%%%%%%%%%%%%%%%%%%%%%%%%%%%%%%%%%%%%%%%%%%%%%%%%%%%%%%%%%%%%%%%%%%%%%%%%%%%%
%%%%%%%%%%%%%%%%%%%%%%%% here comes the title group %%%%%%%%%%%%%%%%%%%%%%%%%%
%%%%%%%%%%%%%%%%%%%%%%%%%%%%%%%%%%%%%%%%%%%%%%%%%%%%%%%%%%%%%%%%%%%%%%%%%%%%%%
\title{Frequency dependent conductivity in the\\ integer quantum Hall effect}
%%%%%%%%%%%%%%%%%%%%%%%%%%%%%%%%%%%%%%%%%%%%%%%%%%%%%%%%%%%%%%%%%%%%%%%%%%%%%%
\author{A. B\"aker and L. Schweitzer} 
%%%%%%%%%%%%%%%%%%%%%%%%%%%%%%%%%%%%%%%%%%%%%%%%%%%%%%%%%%%%%%%%%%%%%%%%%%%%%%
\newcommand{\address}
  {Physikalisch-Technische Bundesanstalt, Bundesallee 100, D-38116
  Braunschweig, Germany}
%%%%%%%%%%%%%%%%%%%%%%%%%%%%%%%%%%%%%%%%%%%%%%%%%%%%%%%%%%%%%%%%%%%%%%%%%%%%%%
\newcommand{\email}{\tt Ludwig.Schweitzer@ptb.de} 
\maketitle
%%%%%%%%%%%%%%%%%%%%%%%%%%%%%%%%%%%%%%%%%%%%%%%%%%%%%%%%%%%%%%%%%%%%%%%%%%%%%
\begin{abstract}
Frequency dependent electronic transport is investigated for a 
two-dimensional  
disordered system in the presence of a strong perpendicular
static magnetic field.  The ac-conductivity is calculated numerically
from  Kubo's linear response theory using a recursive Green's function
technique.  In the tail of the lowest Landau band, we find a linear
frequency dependence for the imaginary part of $\sigma_{xx}(\omega)$
which agrees well with earlier analytical calculations. On the other
hand, the frequency dependence of the real part can not be expressed
by a simple power law.  The broadening of the $\sigma_{xx}$-peak with
frequency in the lowest Landau band is found to exhibit a scaling
relation from which the critical exponent can be extracted.
\end{abstract}
%%%%%%%%%%%%%%%%%%%%%%%%%%%%%%%%%%%%%%%%%%%%%%%%%%%%%%%%%%%%%%%%%%%%%%%%%%%%%

\section{Introduction}
The understanding of the integer quantum Hall effect is intimately related 
to localization and quantum critical phenomena. In a single-particle
description all electronic states are localized, except those at the
critical points near the center of the Landau bands. As a consequence,
at temperature $T=0$ the static longitudinal conductivity
$\sigma_{xx}$ vanishes.  
For finite system size or finite temperature small conductivity peaks
appear at the transition points where the quantized Hall conductivity
changes by $e^2/h$. A further way for obtaining a non-zero $\sigma_{xx}$
is to apply a time dependent electric field and to measure the
frequency dependent conductivity. The broadening of the resultant 
$\sigma_{xx}(\omega)$-peaks with frequency $\omega$ has been detected to
exhibit a scaling relation $\sim \omega^\kappa$ \cite{ESKT93} from
which the product $z_\omega\mu=\kappa^{-1}$ of the dynamical scaling exponent 
$z_\omega$ and the critical exponent $\mu$ governing the divergence of the  
localization length can be obtained. 
However, the frequency-scaling has recently been questioned
experimentally \cite{BMB98} and theoretically \cite{AL96,JZ99}. 

\section{Model and Method}
We consider non-interacting electrons on a square lattice in the 
presence of disorder and a strong perpendicular magnetic field $B$
in a one-band tight-binding approximation. The disorder potentials $w$
are drawn at random from an interval $W/2 \le w \le W/2$ with
constant probability distribution $P(w)=1/W$.  
The Hamilton matrix is 
\begin{eqnarray}
\lefteqn{(H\psi)(x,y) = w(x,y)\,\psi(x,y)+V\,[\psi(x+a,y)+\psi(x-a,y)+}
\nonumber\\
& & \exp(-i2\pi\alpha_Bx/a)\,\psi(x,y+a)+
\exp(i2\pi\alpha_Bx/a)\,\psi(x,y-a)],
\end{eqnarray}
where the lattice constant $a=1$ is taken as the unit of length and 
the transfer term $V=1$ as the unit of energy. 
The magnetic field $B$ is chosen to be commensurate with the
lattice and $\alpha_B=a^2eB/h$ denotes the number of flux quanta per 
plaquette.   

\begin{figure}
\centerline{\resizebox{8cm}{!}{\includegraphics{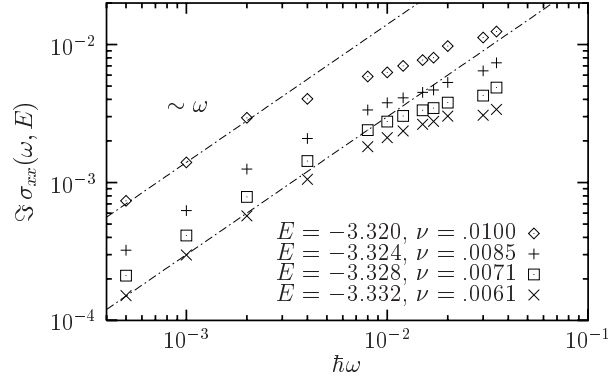}}}
\caption{The imaginary part of the longitudinal conductivity 
$\sigma_{xx}(\omega,E)$ in units $e^2/h$ versus frequency $\omega$. 
For small filling factors $\nu$ a linear frequency dependence is
observed at low $\omega$.}  
\label{Im_xx}
\end{figure}

The Hamiltonian $H^{(N+1)}$ of a lattice system of length $N+1$ can be
split into the Hamiltonian $H^{(N)}_{m,n}$ of the lattice consisting 
of $N$ slices, the part $H_{N+1,N+1}$ of the 
newly added slice, and the coupling $V_{N,N+1}^{} + V_{N+1,N}^{\dag}$.
On this account it is possible to numerically calculate the frequency 
dependent conductivity using a recursive Green's function technique 
developed previously 
\cite{SKM85,Mac85,GB94}, 
\begin{eqnarray}
\lefteqn{\sigma_{xx}(\omega,E_F)=
\lim_{\varepsilon \rightarrow 0^+}\lim_{M \rightarrow \infty}
\frac{e^2}{h\Omega}\frac{1}{\hbar\omega}
\int_{E_F-\hbar\omega}^{E_F}\textrm{Tr}\{(\hbar z)^2 
xG^+_\omega xG^{-}-} 
\nonumber\\
& & (\hbar\omega)^2 xG^+_{\omega}xG^+
+2i\varepsilon x^2(G^+_{\omega}-G^{-})\}\,dE,
\end{eqnarray}
where $z=\omega+2i\varepsilon/\hbar$. $\Omega$ is the system area
$M\times N$, $G^+_\omega=G(E+\hbar\omega+i\varepsilon)$, and 
$G^-=G(E-i\varepsilon)$.
The small imaginary part of the energy 
is necessary for properly taking the thermodynamic limit. It has to
be set to zero, $\varepsilon\to 0^+$, after increasing the 
system width $M$ to infinity, or at least after out-ranging  
an effective system width introduced by the frequency, 
$L_\omega=(1/\rho\hbar\omega)^{1/2}$ ($\rho=$ density of states), which
is relevant near the critical point.
The Green's functions of a system of length $N+1$ can also be
calculated recursively via Dyson's equation with $V^{}_{N,N+1}$ 
as the interaction. 

\begin{figure}
\centerline{\resizebox{8cm}{!}{\includegraphics{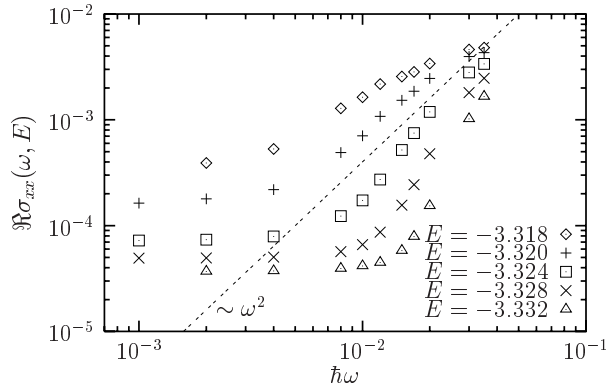}}}
\caption{The real part of the longitudinal conductivity
$\sigma_{xx}(\omega,E)$ in units $e^2/h$ as a function of frequency $\omega$. 
The location of the Fermi-energies $E$ in the tail of the lowest
Landau band correspond to filling factors $\nu=1.2\cdot 10^{-2}$\,($\Diamond$),
$\nu=1.0\cdot 10^{-2}$\,($+$), $\nu=8.5\cdot 10^{-3}$\,($\Box$), 
$\nu=7.1\cdot 10^{-3}$\,($\times$), $\nu=6.1\cdot 10^{-3}$\,($\triangle$). 
The broken line represents a frequency dependence $\sim \omega^2$.}  
\label{Re_xx}
\end{figure}

\section{Results and discussion}
The imaginary and real parts of the frequency dependent longitudinal
conductivity $\sigma_{xx}(\omega,E)$ are shown in Figs.~\ref{Im_xx}
and \ref{Re_xx}, respectively. The Fermi-energy $E$ is situated in the
lowest tail of the disorder broadened Landau band corresponding to small
filling factors $\nu$. The system parameters are taken to be as
follows: width $M=32$, length $N=1.3\cdot 10^5$, the imaginary part of the
energy $\varepsilon=4\cdot 10^{-4}$, disorder strength $W=0.1$, 
and magnetic field $\alpha_B=1/8$. The corresponding cyclotron energy is
$\hbar\omega_c\simeq 0.78$ in our units.
For $\omega$ smaller than a crossover frequency $\tilde{\omega}(E)$, a
linear frequency dependence can be observed. Such a behavior has been
suggested previously for the low frequency limit from analytical
calculations based on a one-instanton approximation \cite{VE90},
\begin{equation}
\Im\, \sigma_{xx}(\omega,E)=-2e^2 l_B \omega\rho(E),
\end{equation}
where $\rho(E)$ is the density of states and $l_B=(eB/\hbar)^{-1/2}$
the magnetic length.
Our data also show that the absolute value of $\Im\,\sigma_{xx}(\omega,E)$
increases with increasing Fermi-energy (filling factor) which is caused
by the rising density of states. 

While the imaginary part of the frequency dependent longitudinal
conductivity is in accord with the analytical predictions, this is not
the case for the real part.  
According to the calculations of Refs.~\cite{VE90,Ape89}, for $E$ in
the lowest localization regime, the real part of $\sigma_{xx}(\omega,E)$  
should obey a quadratic frequency dependence for small frequencies,
\begin{equation}
\Re\, \sigma_{xx}(\omega,E)=c\,\omega^2 \ln(1/\omega^2),
\end{equation}
where $c$ is a constant.
Our results shown in Fig.~\ref{Re_xx} are not compatible with this suggestion.
For $\omega \lesssim \tilde{\omega}(E)$, we observe an almost frequency 
independent conductivity, which for larger $\omega$ is followed by a 
pronounced increase. The steepness of this increase strongly depends
on the position of the Fermi-energy and thus can not be fitted by a 
single power law. Both the exponent and $\tilde{\omega}(E)$, the starting
point of the frequency dependence, increase with 
decreasing filling factor. We have checked that the frequency 
dependence shown in Figs.~\ref{Im_xx} and \ref{Re_xx} is neither
influenced by the choice of $M$ nor $L$. 
We believe that the absence of a single power-law behavior in
$\Re\,\sigma_{xx}(\omega,E)$ is due to the circumstance that the
applied frequency is not small in comparison with the Landau
band width in our calculations. A larger disorder $W$ would improve
the situation.

\begin{figure}

\centerline{%
\resizebox{8cm}{!}{\includegraphics{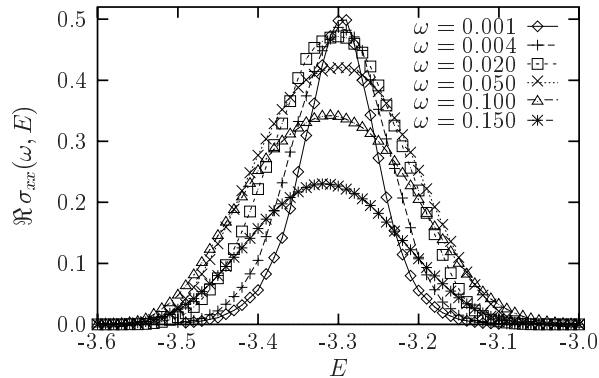}}}
\caption{Frequency broadening of the conductivity peak $\sigma_{xx}(\omega,E)$ 
in units of $e^2/h$ versus Fermi-energy $E$ for $M=32$, $L=10^5$,
$W=1$, $\alpha=1/8$, and $\varepsilon=0.0004$.}  
\label{sxx_E}
\end{figure}

Turning now to the behavior of $\Re\,\sigma_{xx}(\omega,E)$ near
the Landau band center where frequency scaling has been observed in 
experiments \cite{ESKT93}. 
In Fig.~\ref{sxx_E} the broadening of the conductivity peak with frequency
is shown as a function of energy. With increasing $\omega$ the
halfwidth $\Delta_E$ increases while the peak height is reduced. We find a
power law relation $\Delta_E \sim \omega^\kappa$ with $\kappa=0.2\pm 0.01$ 
which is close to $(z_\omega \mu)^{-1}=0.21$ expected from
$z_\omega=2$ (non-interacting electrons) and $\mu=2.35$ \cite{Huc92}.
Thus, in contrast to Refs.~\cite{AL96,JZ99} our results describe a frequency 
scaling similar to what has been observed experimentally in Ref.~\cite{ESKT93}.
While the origin of $z_\omega\approx 1.19$ reported for
non-interacting electrons in Ref.~\cite{AL96} is completely unclear, 
the absence of scaling at low frequencies in Ref.~\cite{JZ99} is attributed 
to the finite bandwidth of delocalized states in their model. Whether
this can serve as an explanation for the experiment \cite{BMB98}
remains to be seen.

%\section{Conclusions}
In conclusions, we have presented results of numerical calculations
for the frequency dependent longitudinal conductivity within the
lowest Landau band. The broadening of the $\sigma_{xx}$-peak exhibits
a power-law behavior in agreement with the notion of frequency scaling.
For small $\omega$, the imaginary part shows a 
$\sigma_{xx}(\omega)\sim \omega$ relation in the lowest Landau band tail 
in accord with suggestions from analytical considerations.

\end{document}